# A new integrating understanding of superconductivity and superfluidity

Jie Han (jiehan@yahoo.com)

**An integrating theoretical scenario of superconductivity and superfluidity has been built. It reduces to the special BCS superconductivity mechanism for conventional superconductor and to a new theory for high transition temperature superconductors, which can explain the recent angle-resolved photoemission experiments and the earlier nuclear spin-lattice relaxation rate experiments. Both experiments suggest the existence of pairing carriers and the normal state energy gap well above the transition temperature of superconductivity. A powerful and workable experiment designed to validate this scenario is also put forward for the experimenter.**

---

BCS theory has been the dominant theory in the research realm of superconductivity mechanism since 1957. The situation did not have any substantial variation until the discovery of high transition temperature cuprate superconductors in 1986. Since then, because of the great difficulties that BCS theory has met in explaining some new experimental results of cuprate superconductors, various theories based on different interaction pairing mechanisms have been put forward in order to obtain such a high transition temperature. However, all these theories met difficulties in explaining some anomalous phenomena, such as null isotope effects, anomalous nuclear spin-lattice relaxation rate, anomalous angle-resolved photoemission spectroscopy (ARPES) etc. In the author's point of view, although these theories were based on interactions other than BCS interaction, they, in fact, were still theories that were developed in the framework of BCS theory. i.e., some interactions lead to the formation of pairing carriers and energy gaps in superconductivity states.

There are enough facts that make the author of this article believe that the BCS pairing mechanism is not always a **necessary and sufficient** condition for the occurrence of superconductivity. Of course, in any case, pairing mechanism should be a **necessary** condition. In fact, as early as 25 years ago, the discovery of the reentrant superconductors (1, 2, 3) appeared to be an early experimental confirmation that BCS pairing mechanism was not a sufficient condition for superconductivity. In those years, a pair breaking theory (induced by magnet ordering) was put forward to explain the re-entrant phenomenon. The great success of BCS theory excludes any possibility of calling into question BCS superconductivity theory. Even though, at the beginning of BCS theory, M. R. Schafroth (4, 5) and C. G. Kuper (6) pointed out that BCS theory could not establish the Meissner effect and the vanishing of the ideal resistance, (when T≠0). The recent discoveries of high transition temperature superconductors (HTS) force scientists to question whether BCS theory needs correction, particularly in the realm of HTS. In this article, a new superconductivity theoretical framework is suggested, which reduces to BCS theory for conventional superconductors and to a wholly new mechanism for HTS that explains significant anomalous phenomena mentioned above. It should be pointed out that the present work concentrates on building a general framework of superconductivity, so that low and high temperature superconductors can be treated in the same theoretical framework without generating so many anomalies. At the same time, pairing mechanism remains an open question, leaving just enough space for further development of superconductivity theory. However, from this simple theoretical framework, one may expect many specific conclusions to explain those anomalous effects and clarification for further development of superconductivity theory.

A long-standing prejudice is that particle pairing transition temperature $T_p$ has always been regarded as the superconductivity transition temperature $T_c$. But, with the application of this new theoretical framework, any previous or later calculation of transition temperature $T_p$ based on interaction pairing mechanism should first be compared with a new characteristic temperature $T_b$ (transition temperature for superfluidity) in order to get the real superconductivity transition temperature $T_c$, which then can be compared with the experimental results of $T_c$. The following will be, first, the suggested theoretical framework that stipulates the relation among $T_b$, $T_p$, $T_c$. Second, an explanation and application of this relation. Third, experimental evidence of it, fourth, a direct experimental design to demonstrate this theoretical framework for the experimenter, and finally a conclusive review and prospect.

**Unified criterion for superconductivity and superfluidity.** It was generally accepted that superconductivity comes from the superfluidity of the charged particles, which leads scientists to regard superfluidity transition temperature $T_b$ (identical with Bose-Einstein condensation temperature) as the critical transition temperature for superconductivity in terms of the work of F. London. But, for element superconductors, C. Kittel (7) pointed out that the Bose condensation temperature calculated for metallic carrier concentrations is of the order of the Fermi temperature ($10^4 \sim 10^5 K$). This rules out any possibility of regarding $T_b$ as $T_c$. Later in this paper it is pointed out that Boson concentrations of element superconductors are not of the order of general metallic carrier concentrations. In fact, the concentrations of Bosons should be $\sim(10^{-5} \sim 10^{-4})$ of metallic carrier concentrations. (8, 9) Kittel's calculation of $T_b$ needs substantial correction.

A critical analysis shows that there exist three significant transition temperatures. Their practical denotations are as: $T_b$, superfluidity transition temperature for Boson or Bosonlike systems, which is identical with Bose-Einstein condensation temperature of the system. $T_p$, the critical transition temperature of forming particle pairs for Fermion systems, which means that, when $T>T_p$, all pairing Fermions will be broken from pairing state. Of course, pairing particle interaction should not be considered as the electron phonon interaction only. Other attractive interactions can also serve as pairing mechanisms; the real $T_p$ is obtained after considering all possible pairing interactions. $T_c$ is the critical transition temperature of superconductivity and superfluidity. For a Boson system, one has, $T_c=T_b$, which means that the superfluidity transition temperature is identical with the Bose-Einstein condensation temperature. For a Bosonlike system formed by pairing Fermions, a formula that stipulates the relations among $T_b$, $T_p$, $T_c$. is provided here: (attention: $T_b$ should be obtained after formation of carrier pairs).

$T_c = $ **Minimum $[T_b, T_p]$** (1)

An appropriate application of the formula above can easily explain many significant experimental results concerning BCS element superconductors and HTS superconductors.

**Explanation and application of Formula (1).**
According to the BCS framework, not all free electrons will become Cooper pairs and therefore become superconducting electrons. The interaction that leads to the formation of Cooper pairs can only act in the energy range of magnitude order $E_g$ (BCS energy gap) near $E_F$ (Fermi energy), which means that the maximum electron number density that can form electron pairs is about $(E_g/E_F)n$, where $n$ is the total free electron number density; at $T_p$, only a small part of the total free electron number density $n$ can form the so-called Copper pairs. In my view, electron pair number density is a constant for $T<T_p$, and at $T=T_p$, only those free electrons with energy above the Fermi energy $E_F(T_p)$ can form Cooper pairs (8). The number of these electrons is:

$$N_F = \int_{E_F}^{\infty} f(E) g(E) dE \qquad (2)$$

This is a general formula. f(E) is the Fermi distribution function, and g(E) is the state density. It is natural to consider a case with $kT_p/E_F(T_p) << 1$ which is very helpful in obtain $N_F$ from equation (2):

$$N_F \approx \frac{3kT_p N}{2E_F(0)}\left[\ln 2 + \frac{\pi^2 kT_p}{24 E_F(0)}\right]$$
$$\approx \frac{3\ln 2}{2}\frac{kT_p N}{E_F(0)} \approx 1.04 \frac{kT_p N}{E_F(0)} \qquad (3)$$

From equation (3), the number density of Cooper pairs is:

$$n_b = \frac{3\ln 2}{4}\frac{kT_p}{E_F(0)} n \qquad (4)$$

Where, $n$ is the number density of free electrons and $n_b$ is the most important quantity for calculation of $T_b$ and $T_c$. It also should be mentioned that these approximations should be reasonable because of the smaller value of $kT_p/E_F(0)$ and the larger value of the denominator of f(E). It is believed, for Boson and Bosonlike systems, the transition temperature of superfluidity is identical with the Bose-Einstein condensation temperature of the system. Because of the difficulties in obtaining Bose-Einstein condensation temperature for interactive Bosonlike systems, We use the Bose-Einstein condensation temperature of ideal Boson systems to replace that of interactive Bosonlike systems and hope to get a reasonable result as in the case of $^4$He. Then, one gets the transition temperature formula for this system:

$$T_b = \frac{2\pi \hbar^2 n_b^{2/3}}{(2.612)^{2/3} k\, m_b} \qquad (5)$$

Where, $\hbar$: Plank constant, $m_b$: Boson mass, $n_b$: Boson number density, $k$: Boltzman constant. We know that:

$$E_F(0) = \frac{\hbar^2}{2m}\left(3\pi^2 n\right)^{2/3} = \frac{\hbar^2}{m_b}\left(3\pi^2 n\right)^{2/3} \qquad (6)$$

Take (4) and (6) into (5), and one finally has the formula.

$$T_b = \frac{2\pi \hbar^{2/3}(0.75\ln 2)^{2/3}}{(2.612)^{2/3}(3\pi^2)^{4/9} k^{1/3}}\frac{n^{2/9}}{m_b^{1/3}} T_p^{2/3} = H\frac{n^{2/9}}{m_b^{1/3}} T_p^{2/3} \qquad (7)$$

Here the constant $H \approx 4.42\times 10^{-16}\, J^{1/3}K^{1/3}S^{2/3}$ (**SI units**).

The following are a few calculated consequences for some typical superconductors.
**Sample one**: BCS element superconductor: Aluminum. We have: $n=18.06\times 10^{28}\, m^{-3}$, $m_b=2m_e=18.22\times 10^{-31}\, k_g$, $T_c= 1.14$ K. Assume that $T_p= T_c$, then, we get $T_b\, (Al)=12.53$ K. (If mass correction induced by strong coupling has been considered, a lower $T_b$ can be obtained.) It's reasonable that one uses $T_c=1.14$ K to replace $T_p$ for Al. because $T_p<T_b$, so we have $T_c=Minimum\ [T_p,T_b]=T_p$. for other element superconductors, one has the same conclusion, i.e., $T_c=Minimum[T_p,T_b]=T_p$. One also should pay attention to the fact that $T_b$ of element superconductors is far lower than $10^4\sim10^5 K$ of Kittel's results (7). More data of $T_b$ and $T_p$ for BCS element superconductors are given in table one.

Table one: $T_p\ (=T_c)$, $T_b$, $n$, $n_b$ for element superconductors (SI Units)

| element | W | Hf | Ir | Ti | Cd | Zr | Ru | Zn | Ga | Mo |
|---|---|---|---|---|---|---|---|---|---|---|
| $n\times10^{-28}$ | 37.08 | 18.08 | 28.24 | 22.64 | 9.28 | 17.16 | 58.88 | 13.10 | 15.30 | 38.52 |
| $n_b\times10^{-23}$ | 0.09 | 0.83 | 1.13 | 2.92 | 3.11 | 3.75 | 5.25 | 5.49 | 7.16 | 8.22 |
| $T_p$ | 0.01 | 0.12 | 0.14 | 0.39 | 0.56 | 0.55 | 0.51 | 0.88 | 1.09 | 0.92 |
| $T_b$ | 0.6 | 2.79 | 3.42 | 6.45 | 6.73 | 7.62 | 9.53 | 9.82 | 11.72 | 12.85 |
| element | Tl | In | Hg | Sn(w) | La | Ta | Pb | V | Tc | Nb |
| $n\times10^{-28}$ | 10.50 | 11.49 | 8.52 | 14.48 | 8.10 | 27.75 | 13.20 | 36.10 | 49.28 | 27.80 |
| $n_b\times10^{-23}$ | 13.85 | 20.31 | 22.43 | 24.00 | 31.89 | 35.90 | 44.97 | 47.06 | 75.40 | 76.17 |
| $T_p$ | 2.39 | 3.40 | 4.15 | 3.72 | 6.00 | 4.48 | 7.19 | 5.38 | 7.77 | 9.50 |
| $T_b$ | 18.20 | 23.48 | 25.10 | 26.25 | 31.73 | 34.33 | 39.90 | 41.12 | 56.31 | 56.69 |

**Sample two**: high $T_c$ cuprate superconductors. For $La_{2-x}Sr_xCuO_4$, $n\approx6\times10^{27}\ m^3$, (10), $m_b=2m_{eff}\approx8m_e=72.88\times10^{-31}$ kg. (11) $T_c=38\ K$, then easily one has, $T_b\approx3.395T_p^{2/3}$, again let $T_p=T_c$, one has $T_b=38.37\ K>T_p$, and $n_b=3.39\times10^{25}\ m^3$, It can be seen that the assumption $T_p=T_c$ is reasonable. But it also should be pointed out that for $La_{2-x}Sr_xCuO_4$, generally one has $T_p\sim T_b\sim T_c$ for different x (i.e., different $n$ and $m_b$). One can get two conditions for it. The first is $T_p\sim T_b$, but $T_p<T_b$. The second is $T_p\sim T_b$, but $T_p>T_b$. For the two conditions above, one can obtain basically the same results for superconductivity transition temperature $T_c$. However, for the isotope effect, one can obtain significant discrepancy because the isotope effect is induced by pairing interaction, thus, for the two conditions above, one has $T_p\propto M^\alpha$ (if applicable); here $\alpha$ is the isotope effect exponent. For the first condition, because $T_p<T_b$, one has $T_c=T_p\propto M^\alpha$. For the second condition, one has $T_p>T_b$, therefore, $T_c=T_b\propto M^{2\alpha/3}$, which gives an isotope exponent $2\alpha/3$. Finally we see that even for the same structure of $La_{2-x}Sr_xCuO_4$, the isotope effect exponent is not a single value. Instead, it has two values, $2\alpha/3$, and $\alpha$, which is in agreement with the experiments of M.L.Cohen et al. (12)

Now, let's see what can be got for $Y_1Ba_2Cu_3O_7$ ($T_c=90K$). One has $n\approx9\times10^{27}\ m^3$. (10,13). But for $m_b$, there is a large selection scope, from $m_b\approx6m_e$ to $m_b\approx20m_e$. It is considered that $m_b\approx6m_e$ is a reasonable one (10, 14). Then one gets $T_b=4.089T_p^{2/3}$, one easily sees that, for $T_p>68.35\ K$, $T_b<T_p$. Now, it is known that $T_p\geq T_c=90\ K$, so $T_b<T_p$. Thus, for $Y_1Ba_2Cu_3O_7$, in any case one always has $T_c=T_b<T_p$. This is a significant discrepancy from conventional element superconductors. Here the isotope effect exponent is $2\alpha/3$ (if $T_p\propto M^\alpha$). Finally, for $Y_1Ba_2Cu_3O_7$, $T_c=T_b=90K$, $T_p=103.27K$, $n_b=7.92\times10^{25}\ m^3$. What follows are the significant facts that support this larger $T_p=103.27\ K$, and especially explain the anomaly of nuclear spin-lattice relaxation rate for superconductor $Y_1Ba_2Cu_3O_{6.67}$ with $T_c=60\ K$. For this sample, $n\approx8\times10^{27}\ m^{-3}$, but for this superconducting phase, it is considered that a larger effective mass (15) $m_b\approx18m_e$ should be used. Then $T_b=2.762T_p^{2/3}$, and again, one finds $T_b<T_p$, if $T_p>22.07\ K$. Finally we have $T_b=T_c=60\ K$, $T_p=101.26\ K$, $n_b=2.24\times10^{26}\ m^3$. Now, let's see what one can say for nuclear spin-lattice relaxation experiments of $Y_1Ba_2Cu_3O_{7-\delta}$ with superconducting transition temperature of *60 K* and *90 K respectively*. For experimental results, see W. W. Warren et al. (16) and M. Takigawa et al. (17). It is known that when $T_p$ is arrived, because of the formation of carrier pairs, the state of conducting carrier will produce a deep change, which definitely, will cause the corresponding change of nuclear spin-lattice relaxation rate $(T_1T)^{-1}$. For $La_{2-x}Sr_xCuO_4$ and $Y_1Ba_2Cu_3O_7$, although $T_p$ may not be $T_c$, for the sake of $T_p\sim T_b\sim T_c$, the anomaly of curve $(T_1T)^{-1}$ vs. $T$ is not so obvious. Nevertheless, for *60 K* $Y_1Ba_2Cu_3O_{6.67}$ superconductor, because $T_c=60\ K$, $T_p=101.26\ K$. Curve $(T_1T)^{-1}$ vs. $T$ will produce a distinct anomaly at $T_p=101.26\ K$, instead of at $T_c=60\ K$. Here, the nuclear spin-lattice relaxation rate experiment is easily explained in this new superconductivity theoretical framework without any anomaly. This provides substantial support for this theoretical framework.

The recent ARPES experiments (18, 19) reveal evidence of an energy gap in the normal state excitation spectrum of the underdoped cuprate superconductor $Bi_2Sr_2CaCu_2O_{8+\delta}$, and a sharp peak feature of superconducting state for underdoped or overdoped samples. These contradict previous theories. But, with the application of the new scenario mentioned in this paper, this puzzle is explained. In fact, underdoped samples generally correspond to the case: $T_c=T_b<T_p$, which means that carrier pairs and energy gaps preform before the occurrence of superconductivity. That is, carrier pairs form without long-range coherence well above the superconducting realm. In the mean time, overdoped samples generally correspond to the case: $T_c=T_p<T_b$, (i.e., standard BCS case) which means that when energy gap and carrier pairs are formed, the system will go into the superconducting state automatically. By the way, the optimally doped $T_c$ is obtained when $T_b\sim T_p$. As for the sharp peak feature of superconducting state, this paper suggests that it might be the natural consequence of Bose-Einstein condensation. It even is hoped that further study of the peak intensity evolution with temperature will reveal quantitative evidence of Bose-Einstein condensation. Qualitatively, when temperature increases from 0 $K$ to $T_c$, the sharp peak will go down from the maximum intensity to that of normal state.

**The difference between superconducting electron pairs and general carrier pairs.**
At temperature $T_p$, although carrier pairs are formed with the concentration $n_b$, it doesn't mean that these carrier pairs become superconducting electron pairs. In fact, at temperature $T$ ($T<T_p$, $T_b$), the number density of superconducting electron pairs $n_S$, is given by $n_S/n_b=1-(T/T_b)^{3/2}$. For the case: $T<T_p<T_b$, when $T$ increases from zero to $T_p$, $n_S$ tends to zero because of broken pairs and superconductivity disappears. For the case $T<T_b<T_p$, when T increases from zero to $T_b$, $n_S$ tends to zero and superconductivity disappears even though carrier pairs still exist.

**Experiment designed to verify this integrating scenario.**
The thought to consider superconductivity as a result of superfluidity of charged carrier pairs has been suggested in the early years of superconductivity theory. But according to the scenario outlined above, it can be seen that $T_c$ and $T_b$ are not always the same temperature, sometimes, $T_c = T_p$, as in the case of BCS conventional superconductors, and sometimes, $T_c = T_b$, as in the underdoped HTS cuprate superconductors. These two competing transition temperatures make understanding the mechanism of superconductivity even more complex. However, the great success of BCS theory cannot deny that the superfluidity of the charged carrier pairs is a control factor for the occurrence of superconductivity. Therefore, a workable and decisive experimental design is suggested for the experimenters to validate the viewpoint presented in this paper. The experimental principle is simple. Because the most direct proof of the pairing effect of charged carriers is the Giaever effect, this paper suggests a new Giaever sandwich like Al/Al$_2$O$_3$/HTS. Where HTS may be the underdoped $Y_1Ba_2Cu_3O_{6.67}$, or Bi$_2$Sr$_2$CaCu$_2$O$_{8+\delta}$ (both should have $T_c\sim60$ $K$) it is suggested that the similar Giaever effect can be seen in the temperature range T~(*70~80 K)*. Because the temperature range chosen is just in the range $T_b=T_c<T<T_p$. this experiment should be an executable one under current experimental conditions, though it can be a difficult one due to the very reactive surface of underdoped samples. A further consideration may be the new Josephson sandwich formed by the underdoped cuprate superconductors mentioned above. Because of the conventional Josephson effect observed under the condition $T < T_c = T_p <T_b$, it is reasonable to expect some new consequences from this new Josephson sandwich in the temperature range $T_c = T_b< T < T_p$. Of course, it will deepen the current understanding of the pair tunneling effect and the coherence effect among carrier pairs. They are combined in the case of conventional superconductors and are difficult to distinguish from one another. The third consideration is a low temperature specific heat experiment at appropriate temperature.

**Conclusion, discussion and prospect of this new scenario.**
Based on the essential understanding that there should exist a unified theory for the conventional BCS superconductors and the high transition temperature cuprate superconductors, a new theory framework is put forward, which reduces to BCS pairing theory for convention superconductors. However, for cuprate superconductors, it develops into a very new theory that suggests the origin of the anomalies in the normal state of cuprate superconductors is the same as that which induces Cooper's pairing effect of superconducting states. This viewpoint is in agreement with many other theories, such as RVB theory, Luttinger liquid theory, marginal Fermi liquid theory, etc. From this new theory framework, many significant conclusions can be drawn.

(a) Superconductivity transition temperature depends on the delicate balance among some competing factors from which pairing effect and Bose-Einstein condensation effect are two important factors. Which factor becomes the control or dominant one depends on the samples (doping) and it is this sample dependence that makes the research on the mechanism of superconductivity more complex.
(b) Because this theoretical framework stands above pairing theory and Bose-Einstein theory, the properties of the superconducting state, e.g., zero resistance and Meissner effect, will result from this theory unambiguously.
(c) From formulae (1) and (4), one sees that a larger $T_c$ requires a larger $T_p$ and $T_b$, the cuprate superconductors are samples with larger $T_p$, and as for a larger $T_b$, one needs a larger $n_b$, which requires a smaller $E_F(0)$. A smaller $E_F(0)$ means a larger effective mass, which means heavy Fermions. This theory appears to reveal its potential in application to heavy Fermion superconductors.
(d) Many people know that element superconductors with lower carrier concentrations $n$ have higher transition temperatures $T_c$, which appears difficult to understand. In fact, these superconductors generally have $T_p$ and $E_F(0)$ such that the ratio $T_p/E_F(0)$ has a larger value. Therefore, a larger effective carrier pair concentration $n_b$ can be obtained even though it has a smaller $n$.
(e) Unusual energy gap. We know that for $Y_1Ba_2Cu_3O_7$ ($T_c\sim90$ $K$), the experimental value of the energy gap is $2\Delta_{ab}(0)/(KT_c)=6\sim8$, where, $2\Delta_{ab}(0)$ is energy gap at temperature $T=0$ $K$. According to the scenario in this paper and BCS theory, it is found that the significant quantity is $2\Delta_{ab}(0)/kT_p$, rather than $2\Delta_{ab}(0)/kT_c$, because $T_c<T_p$, for Y$_1$Ba$_2$Cu$_3$O$_7$, one can get a smaller value of the quantity $2\Delta_{ab}(0)/kT_p$. This finally clarifies the so-called unusual energy gap phenomenon. i.e., $E_g$ induced by pairing effect scales with $T_p$ instead of $T_c$, which solves this question. In fact, ARPES experiments suggest the lack of scaling of the energy gap with $T_c$.
(f) From formulae (4) and (7), one understands that $T_p$ and $T_b$ both depend on $n_b$ and $n$ sensitively, which strongly implies the possibility of the application of high pressure effect on the enhancement of $T_c$. This is especially true for the cases for which $T_c=T_b<T_p$, such as HTS superconductors. On the other hand, the high-pressure effect will deepen the current understanding of the superconductivity mechanism.
(g) Because the low temperature specific heat of the ideal Boson is $\sim T^{3/2}$ (when $T < T_b$), which is different from that of the free electron ($\sim T$), then if carrier pairs are considered as Bosonlike particles, further measurement of specific heat will give meaningful results.
(h) It is believed that the curious properties of the superconducting state come from the corresponding properties of carrier pairs, i.e., carrier pairs present themselves as a Bose-Fermi statistical duality, which will merit further study.

The author believes that this theoretical framework clarifies many key problems in the realm of high temperature superconductors and this new scenario will stimulate further research in this field, especially in the pursuit of new $T_p$ that originates from interactions other than electron-phonon interaction. Otherwise, in the case of underdoped superconductors, $T_p$ may not be $T_c$, even when one has a correct $T_p$.

**Acknowledgements:** The author hopes to thank Dr. Claude Jacques, Dr. George Chapman of NRC for help in preparing this paper, Professor A. M. Tremblay of The University of Sherbrooke for instructive and stimulating discussion.